\documentclass[10pt,conference]{ieeeconf}
\usepackage{verbatim}
\usepackage{amsmath}
\usepackage{amssymb}
\newtheorem{thm}{Theorem}
\newtheorem{lem}{Lemma}
\newtheorem{defn}{Definition}
\newtheorem{rem}{Remark}
\newtheorem{example}{Example}
\newtheorem{cor}{Corollary}
\newtheorem{ass}{Assumption}
\IEEEoverridecommandlockouts  
\usepackage{color}

\begin{document}
 \title{A Scale-Space Theory of Memory} 

\author{
\authorblockN{
Dongjun Wu\authorrefmark{1},
Rodolphe Sepulchre\authorrefmark{1}\authorrefmark{2}
}
\authorblockA{
\authorrefmark{1}
		Department of Electrical Engineering, ESAT-STADIUS,
		KU Leuven, Leuven, Belgium \\
\authorrefmark{2}
        Department of Engineering, 
        University of Cambridge, Cambridge, United Kingdom 
}
}

\maketitle
\begin{abstract}
Scale-space theory has emerged as a central attempt
to restrict the {\it general} theory of signal processing 
with geometric constraints inspired from the animal world. 
In the mathematical language of operator theory, 
scale-space theory restricts the Toeplitz operator. 
The present paper investigates scale-space theory for
memory systems. We investigate how scale space theory
restricts a memory system encoding past information, that is,
restricts the Hankel operator. Our main result is to establish a close correspondence
between scale-space theory and the long-studied {\it relaxation} property.
\end{abstract}

\section{Introduction }\label{sec:introduction}

The question of \emph{memory} has lost much of its importance in
modern system theory, owing to the concept of {\it state}: whenever
a dynamical system is equipped with a state-space representation, its
state {\it is} the memory. This mathematical abstraction is powerful,  
but perhaps too general for engineering purposes. What singles out a {\it physical}
or {\it physiological} memory system from merely having a state-space representation?

Such questions were central in the Volterra-Wiener
tradition of nonlinear systems theory. The key modeling question of
an amplifier was to model its {\it memory}. Both Volterra and Wiener already
recognized the decaying influence of the remote past as a fundamental
modeling principle \cite{Volterra1959,Wiener1958}. 
In continuum mechanics, Coleman and
Noll formalized \emph{fading memory}
through continuity of constitutive functionals in weighted history
spaces, which discount the remote past \cite{ColemanNoll1960,ColemanNoll1961}. Coleman
subsequently developed a thermodynamic framework for materials with
memory \cite{Coleman1964}. In nonlinear systems theory,
Boyd and Chua formulated fading memory for causal time-invariant operators
and established its crucial role in nonlinear system approximation
\cite{boyd1985fading}. In that input-output approach, fading memory 
was more than a  mere parametrisation of the past. It endowed the 
system with key input-output properties that differentiate
mathematical modeling from physical modeling. For instance, a striking
side result in 
\cite{boyd1985fading} was the observation that a causal LTI operator
has the convolution property if and only if it has fading memory, see also \cite{SepulchreFading2021}. 

For nonlinear systems, the recent article \cite{Bainier2026TAC}
clarifies the input-output stability properties of a state-space model equipped
{\it with} fading memory.

The objective of the present article is to further restrict the concept
of memory, with the view that fading memory is still too general to capture the design constraints of an \emph{efficient} memory machine.
We attack this general
question by mimicking the angle of scale space theory.

Developed extensively in computer vision and signal processing, notably by Lindeberg \cite{lindeberg1994},
scale-space theory addresses how a signal should be represented at different resolutions
without introducing artifacts. Two principles are particularly relevant.
The first is \emph{non-creation of feature}s. That is, the number
of local extrema or zero crossings should not increase from finer
to coarser representations \cite{Lindeberg2016}. The second is \emph{scale-covariance},
in the sense that a dilation of the input should correspond to a similar
dilation of its output, together with an appropriate change of the
scale parameter \cite{Lindeberg2023}. These principles constrain
how the structure of a signal is represented and organized across
scales.

The two principles are natural for defining memory. A memory operator
should not depend on an arbitrary choice of time unit or observation
scale -- stretching the temporal axis should correspond only to an
appropriate change of the memory scale. This motivates scale-covariance.
Moreover, future information should not create new temporal features
that were absent from the input history. This motivates a \emph{variation-diminishing}
principle. Thus, scale-covariance describes how memory is organized
across temporal scales, while variation-diminishing constrains how
information from the past may be carried into the future.

Scale-space theory is, in essence, a filtering theory concerned with
extracting features of a signal at different scales \cite{lindeberg1994,Lindeberg2023}.
In the causal temporal setting, signal processing is concerned with mapping the \emph{future to the future}. A memory, on the other hand, is concerned with what a dynamical system retains from past inputs and how these retained effects determine
its subsequent response. Thus it is concerned with mapping the \emph{past to the future}. This distinction has substantial mathematical consequences.
In particular, the property of non-creation of new features will lead
to rather different constructions, and the scale-space theory of signal
processing must be adapted to characterize memory operators.
This motivates a scale-space theory tailored to memory operators that
we investigate in this paper. 

The main result of the present article is to establish a close correspondence
between scale-space theory applied to memory systems and the system theoretic
concept of {\it relaxation}.

The concept of \emph{relaxation system} has deep roots in physics, particularly
in the study of material response. A material ``remembers'' when its
present response depends not only on its present deformation, but
also on the loading history through which that deformation was reached.
Maxwell's theory of stress relaxation provides an early and enduring
example \cite{Maxwell1867}. In its standard setting, the Maxwell
model consists of an elastic spring in series with a viscous dashpot
and exhibits exponential stress relaxation when strain is held fixed
\cite{Roylance2001}. Such relaxation elements provide elementary
building blocks for models of viscoelastic materials with memory,
whose internal dynamical states carry the effects of past loading
into the subsequent responses \cite{Roylance2001}.

 The physical property of relaxation received a system-theoretic
 treatment in dissipativity theory \cite{Willems1972}. Willems showed that
relaxation systems have completely monotone impulse responses and
admit positive parallel realizations by first-order
lags \cite{Willems1972,Chaffey2023}. However, an important aspect of memory called temporal scale covariance -- which will be discussed later -- is missing from both \cite{Willems1972,Chaffey2023}.

The specific {\bf contributions} of the present article are twofold:
\begin{enumerate}
\item We introduce a scale-space theory for memory operators, 
that relate to the classical concept of relaxation systems of Willems for LTI systems. 
\item By incorporating fading memory as an additional regularity condition,
we establish a universal approximation theorem for nonlinear relaxation
memory.
\end{enumerate}

\section{Hankel operators}

A dynamical system carries information from its past into its future.
Imagine observing the system at the present time, which we take to
be $t=0$. The system has been driven by an input over the past $t\leq0$;
at $t=0$, the input is switched off, and we observe the subsequent
free response for $t\ge0$. Whatever appears in this future response
must have been retained from the past. This motivates taking a \emph{past-to-future
operator} as the basic object for studying memory. 

Throughout this note, signals are scalar-valued. A signal defined
on $\mathbb{R}_{-}=\{x\le0\}$ will be called a \emph{past input},
and a signal defined on $\mathbb{R}_{+}=\{x\ge0\}$ will be called
a \emph{future output}. We take the signal spaces to be 
$L_{2}(\mathbb{R}_{-})$ and $L_{2}(\mathbb{R}_{+})\cap C_{b}(\mathbb{R}_{+})$
respectively, where $C_{b}(\mathbb{R}_{+})$ is the set of bounded
continuous functions on $\mathbb{R}_{+}$. The past input space is
furnished with the usual $L_{2}$-norm, while the future-output space
is equipped with $\|y\|=\|y\|_{L_{2}}+\|y\|_{\infty}$. Both are Banach
spaces. The additional continuity required of the output reflect the
smoothing property of the operators considered below.

We begin with the past input to future output operator
\[
H:L_{2}(\mathbb{R}_{-})\to L_{2}(\mathbb{R}_{+})\cap C_{b}(\mathbb{R}_{+}).
\]
An arbitrary map $H$, however, need not admit a dynamical system
interpretation. If it truly describes the past-to-future behavior
of a dynamical system, then it should be obtainable by restricting
an ordinary whole-line input-output operator to an experiment in which
the future input is set to zero. 

To make this precise, let $E_{-}:L_{2}(\mathbb{R}_{-})\to L_{2}(\mathbb{R})$
denote the zero extension
\[
E_{-}u(t)=\begin{cases}
u(t), & t\le0,\\
0, & t>0
\end{cases}
\]
and let $P_{+}$ and $P_{-}$ denote the restriction to $\mathbb{R}_{+}$
and $\mathbb{R}_{-}$ respectively. 
{} We ask, whether there exists a \emph{unique} operator 
\begin{equation}
F:L_{2}(\mathbb{R})\to L_{2}(\mathbb{R})\cap C_{b}(\mathbb{R})\label{eq:R to R}
\end{equation}
such that $H=P_{+}FE_{-}$.

To make the operator $F$ into a system, we impose two standard requirements,
namely, \emph{causality} and \emph{time-invariance}. $F$ is said
to be \emph{causal} if $u(s)=v(s)$ for all $s\le t$ implies $Fu(t)=Fv(t)$;
it is said to be \emph{time-invariant} if $T_{s}F=FT_{s}$ for all
$s$, where $T_{s}$ is the time-shift operator on on $L_{2}(\mathbb{R})$:
$T_{s}u(t)=u(t-s)$. The two notions are standard, see e.g., \cite{boyd1985fading}.

Causality and time-invariance impose a simple but important consistency
condition directly on the past-to-future map $H$. Suppose a past
input $u$ has generated the future response $Hu$. If we wait for
a time $t>0$, while keeping the input equal to zero, then two descriptions
of the system must agree. We may either advance the already generated
future response by $t$, or regard the original history as having
aged by $t$ and recompute the future from the new present. These
two procedures must produce exactly the same trajectory.

To formalize this consistency, define, for $t\ge0$, the \emph{aging
operator}
\[
A_{t}=P_{-}T_{-t}E_{-}:L_{2}(\mathbb{R}_{-})\to L_{2}(\mathbb{R}_{-}).
\]
Thus $A_{t}u$ is the past history seen $t$ seconds later after the
input has been switched off. On the output side, define the \emph{future
advance operator}
\[
S_{t}y(r)=y(r+t),\quad r>0.
\]
Time consistency then requires
\begin{equation}
HA_{t}=S_{t}H,\quad\forall t\ge0.\label{eq:H-commute}
\end{equation}

Before stating the extension result, we impose a mild regularity condition.
\begin{ass} \label{ass:1}
The past-to-future operator $H$ is continuous
and satisfies
\[
\forall r>0,\quad\sup_{\|u\|_{L_{2}(\mathbb{R}_{-})}\le r}|H(u)(0)|<\infty.
\]
\end{ass}

The preceding conditions, together with an integrability
requirement ensuring that the reconstructed whole-line output belongs
to $L_{2}(\mathbb{R})$, completely characterize those past-to-future
maps that arise from causal time-invariant systems.
\begin{lem}
\label{lem:extension}Let $H$ be a past-to-future operator satisfying
Assumption \ref{ass:1}. Then $H$ has a unique extension to a causal, time-invariant
operator \eqref{eq:R to R} such that $H=P_{+}FE_{-}$ if and only
if \eqref{eq:H-commute} holds and 
\begin{equation}
t\mapsto HP_{-}T_{-t}u(0)\in L_{2}(\mathbb{R}),\quad\forall u\in L_{2}(\mathbb{R}).\label{eq:H-L2}
\end{equation}
In this case the extension is necessarily
\[
Fu(t)=HP_{-}T_{-t}u(0),\quad t\in\mathbb{R}.
\]
\end{lem}

The proof of the lemma is omitted.

We shall henceforth restrict attention to past-to-future operators
that admit the unique causal time-invariant extension characterized
by Lemma \ref{lem:extension}. We call such operators \emph{Hankel
operators}, even if the classical definition of a Hankel operator also assumes linearity. These operators will serve as the concrete mathematical
objects for the loose term \emph{memory operators} in this paper. 

A Hankel operator decomposes the dynamics of a system into two parts.
The past input first drives the system to its present state. That
state summarizes precisely the information from the past that remains
relevant for the future. Once the input is switched off, the state
then generates the future free response:
\[
\text{past history}\;\longrightarrow\text{present state}\;\longrightarrow\;\text{free future evolution}.
\]
Thus a Hankel operator can be viewed as an external representation
of what the system has retained from its past.

For a linear time-invariant system, this picture reduces to the classical
Hankel factorization $H=\mathcal{O}\mathcal{C}$, where $\mathcal{C}$
maps the past input to the state at the present time and $\mathcal{O}$
maps that state to its zero-input future output. In the nonlinear
setting, the same interpretation exists, where the central realization
problem is to identify a \emph{state representation} that contains
exactly the information from the past needed to determine the future
free response \cite{Nerode1958,KalmanFalbArbib1969}.

If we view a Hankel operator as an encoder (i.e., a memory operator) of the past inputs, then $H$ should encode information in a reasonable way.
The introduction of internal states and the associated decomposition mentioned above may obscure, or fail to preserve, the relevant past-to-future behavioral properties of an encoder.
This leads to the central question of this paper: what structural properties should a Hankel operator satisfy to constitute a meaningful representation of memory?

\section{Scale-Space Memory}

\subsection{Scale-covariance and Variation Diminishing}

Two requirements will play a central role. The first concerns \emph{time
scales}. Memory is inherently temporal, but there is generally no
privileged absolute unit of time at which a temporal pattern must
be represented. The same pattern may unfold rapidly or slowly, while
its temporal organization remains unchanged. A memory mechanism should
therefore transform consistently when time itself is rescaled.

This observation leads us to consider not a single Hankel operator,
but a one-parameter family $\{H_{\tau}\}_{\tau>0}$, where the parameter
$\tau$ specifies the memory scale. Small values of $\tau$ corresponds
to fine temporal scales, whereas large values correspond to coarse
scales. Changing the speed of the input should then be accompanied
by the same change in the memory scale.

For $c>0$, define the time dilation operators on past inputs and
future outputs by
\begin{align*}
D^{-}_{c}u(t) & :=u(t/c),\quad t\le0,\\
D^{+}_{c}y(t) & :=y(t/c),\quad t\ge0.
\end{align*}
We can now state the first structural requirement.
\begin{defn}[Scale-Covariance]
\label{def:covariance}The one-parameter family of Hankel operators
$\{H_{\tau}\}_{\tau>0}$ is said to be \emph{scale-covariant} if 
\[
H_{c\tau}D^{-}_{c}=D^{+}_{c}H_{\tau},\quad\forall c>0,\;\forall\tau>0.
\]
\end{defn}

Scale covariance tells us how memory should behave when the temporal
scale is changed. It does not yet tell us what information a memory
operator is allowed to create. This leads to the second requirement.

A memory may forget parts of that past, merge nearby temporal features,
or attenuate them as they recede in time. What it should not do is
manufacture temporal features that were not present in the signal
being remembered. For scalar signals, a particularly natural way of
expressing this principle is through their sign variations \cite{Karlin1968I}.

Consider an ordered sequence of real numbers $(z_{1},\cdots,z_{m})$.
Let $S(z_{1},\cdots,z_{m})$ be the number of sign changes in the
sequence obtained by deleting all zero entries. For a continuous scalar-valued
function $f$ with finite zero on an interval $I$ (in our setting,
either $I=(-\infty,0]$ or $I=[0,\infty)$), define the sign changes
of $f$ as 
\[
S(f)=\sup_{t_{1}<\cdots<t_{m},\,\{t_{i}\}^{m}_{1}\subset I}S(f(t_{1}),\cdots,f(t_{m})).
\]
The value $S(f)$ measures the oscillatory sign structure of the signal,
independently of the locations of its zero crossings.

We require a memory operator not to increase this quantity. Moreover,
when no sign variation is lost, the temporal order of the signs should
be preserved. This is the order-preserving variation diminishing property.

For $u$ defined on $\mathbb{R}_{+}$ or $\mathbb{R}_{-}$, denote
the time reversal of $u$ as $\widetilde{u}(t)=u(-t)$.
\begin{defn}[Order-preserving Variation Diminishing]
\label{def:OVD}Let $X$ be a subset of $L_{2}(\mathbb{R}_{-})\cap C(\mathbb{R}_{-})$.
A Hankel operator $H$ is said to be \emph{order-preserving variation
diminishing} on $X$, written $H\in{\rm OVD}(X)$ if, for every $u\in X$,
with $S(\widetilde{u})<\infty$,
\[
S(Hu)\le S(\widetilde{u})
\]
and whenever equality holds, $\widetilde{u}$ and $Hu$ have the same
ordered sign patterns. 

When $X=L_{2}(\mathbb{R}_{-})\cap C(\mathbb{R}_{-})$, we simply say
that $H$ is order-preserving variation diminishing and write $H\in{\rm OVD}$. 

\end{defn}
\begin{rem}
The role of the set $X$ is important. Variation diminishing may hold
on a restricted class of inputs without holding on the entire signal
space, and different choices of $X$ may therefore lead to different
structural characterizations of the operator; see, for example, \cite{grussler2022variation}. 
\end{rem}

We can now combine the two principles.
There is one additional consistency requirement. That is, if the
memory scale is changed, the corresponding class of admissible signals
must be rescaled with it. This leads to the central definition:
\begin{defn}[Scale-Space Memory] \label{def:ss-memory}
 Let $\{H_{\tau}\}_{\tau>0}$ be a one-parameter family of Hankel
operators, and let $K_{\tau}\subset L_{2}(\mathbb{R}_{-})\cap C(\mathbb{R}_{-})$
be a corresponding family of input sets. The pair $\{H_{\tau},K_{\tau}\}_{\tau>0}$
is called \emph{a scale space memory} if
\begin{enumerate}
    \item the family $\{H_{\tau}\}_{\tau>0}$ is scale-covariant and that $H_{\tau}\in{\rm OVD}(K_{\tau})$, $\forall\tau>0$;
    \item the input sets $\{K_{\tau}\}_{\tau>0}$ are scale-covariant in the sense that $K_{c\tau}=D^{-}_{c}K_{\tau}$ for every $c>0$, $\tau>0$.
\end{enumerate}
When $K_{\tau}=L_{2}(\mathbb{R}_{-})\cap C(\mathbb{R}_{-})$ for every
$\tau$, we simply say that $\{H_{\tau}\}_{\tau>0}$ is a memory scale
space. 
\end{defn}

With this definition at hand, we are ready to characterize scale-space memory rigorously. We start with linear theory.

\subsection{Linear Scale-Space Memory and Relaxation Systems}

The definition of a scale-space memory in the previous
subsection is axiomatic. It does not prescribe a particular
realization. A natural question is therefore how much freedom remains
once these axioms are imposed.

In the linear case, the answer is remarkably rigid. Every linear 
scale-space memory is built from the simplest possible fading-memory
mechanisms: first-order lags. The only freedom lies in how these lags
are combined.
\begin{thm}
\label{thm:lin}$\{H_{\tau}\}_{\tau>0}$ is a linear scale-space memory
\emph{if and only if}, for every $\tau>0$, the unique causal
time-invariant extension of $H_{\tau}$ has the following frequency
domain representation
\begin{equation}
\widehat{F}_{\tau}(s)=\int_{(0,\infty)}\frac{a}{\tau s+a}d\mu(a),\quad\forall\tau>0,\label{eq:H int}
\end{equation}
where $\mu$ is a finite nonnegative Borel measure on $(0,\infty)$
independent of $\tau$, and the corresponding impulse response of
$F_{\tau}$ belongs to $L_{2}(\mathbb{R}_{+})$ for every $\tau>0$. 

Moreover, the measure $\mu$ is uniquely determined by the 
scale-space memory.
\end{thm}

\begin{rem}
Let $g_{\tau}$ be the impulse response of $F_{\tau}$, then
\begin{equation} \label{eq:g_tau}
g_{\tau}(t)=\int_{(0,\infty)}\frac{a}{\tau}e^{-at/\tau}d\mu(a),\quad t\ge0.
\end{equation}
Hence 
\[
\|g_{\tau}\|_{L_{1}}=\mu((0,\infty))<\infty.
\]
Consequently, 
\[
\|F_{\tau}u\|_{\infty}\le\mu((0,\infty))\|u\|_{\infty},
\]
so the system is BIBO stable. Notice that the gain is independent
of $\tau$. Moreover, since $g_{\tau}\in L_{2}(\mathbb{R}_{+})$,
their free responses satisfy 
\[
|H_{\tau}u(t)|\leq\left(\int^{\infty}_{t}|g_{\tau}(r)|^{2}dr\right)^{1/2}\|u\|_{L_{2}}\to0
\]
as $t\to\infty$, for every fixed $\tau>0$ and $u\in L_{2}(\mathbb{R}_{-})$,
exhibiting the relaxation behaviour of the Hankel operators. 
\end{rem}

\begin{rem}
The condition that $g_{\tau}\in L_{2}(\mathbb{R}_{+})$ can be equivalently
characterized by
\[
\int_{(0,\infty)^{2}}\frac{ab}{a+b}d\mu(a)d\mu(b)<\infty.
\]
\end{rem}
The representation \eqref{eq:H int} gives a concrete picture of linear
memory. For each $a>0$, the factor
\[
\frac{a}{\tau s+a}=\frac{1}{(\tau/a)s+1}
\]
is a first-order lag with time constant $\tau/a$. Hence the theorem
states that every linear scale-space memory is a nonnegative
parallel connection of first-order memories. 
The parallel interconnection of first-order lags is
classical and appears in Willems' characterization of relaxation
systems~\cite{Willems1972}. The distinguishing feature of the our characterization is that these lag elements are organized across
temporal scales: their time constants are dilated proportionally
with $\tau$, while their relative weights and normalized rates
remain invariant. This scale organization is encoded by a single
nonnegative measure $\mu$, independent of $\tau$.

\begin{example}[Heat equation]
Consider the heat equation
\begin{equation}
\tau\frac{\partial\xi}{\partial t}(x,t)=\frac{\partial^{2}\xi}{\partial x^{2}}(x,t)+b(x)u(t)\label{eq:Heat}
\end{equation}
on the open interval $(0,\pi)$ with Dirichlet boundary conditions
\[
\xi(0,t)=\xi(\pi,t)=0
\]
and output 
\begin{equation}
y(t)=\int^{\pi}_{0}b(x)\xi(x,t)dx.\label{eq:heat-out}
\end{equation}
Assume $b\in L^{2}(0,\pi)$. The transfer function from $u$ to $y$
is
\[
\widehat{H}_{\tau}(s)=\sum^{\infty}_{i=1}\frac{b^{2}_{n}}{n^{2}}\frac{n^{2}}{\tau s+n^{2}}
\]
in which $b_{n}=\left\langle b,\phi_{n}\right\rangle $. This is exactly
of the form \eqref{eq:H int}, with 
\[
\mu=\sum^{\infty}_{n=1}\frac{b^{2}_{n}}{n^{2}}\delta_{n^{2}}
\]
being a finite discrete measure.
\end{example}

The finite-rank case reduces the distributed spectrum to finitely
many atoms as the following corollary shows.
\begin{cor}
\label{cor:lin} A finite dimensional linear scale-space memory $\{H_{\tau}\}_{\tau>0}$ has the relaxation property, that is, 
the following statements are equivalent
\begin{enumerate}
    \item for some $\tau>0$ 
\[
{\rm rank}H_{\tau}=n<\infty;
\]
\item 
the measure $\mu$ in Theorem \ref{thm:lin} has $n$ atoms,
\[
\mu=\sum^{n}_{i=1}w_{i}\delta_{a_{i}},\quad w_{i}>0,\quad0<a_{1}<a_{2}<\cdots<a_{n};
\]

\item for every $\tau>0$, 
\begin{equation}
\widehat{H}_{\tau}(s)=\sum^{n}_{i=1}\frac{w_{i}}{\tau_{i}s+1},\quad\tau_{i}=\frac{\tau}{a_{i}}.\label{eq:finite H}
\end{equation}
\end{enumerate}
In this case, ${\rm rank}H_{\tau}=n$ for every $\tau>0$.
\end{cor}

\subsection{Scale-Space Theory for Signal Processing}\label{subsec:comparison-signal-processing}

Theorem~\ref{thm:lin} and Corollary~\ref{cor:lin} allow a concrete
comparison with temporal scale-space theory \cite{Lindeberg2016}.
Both constructions use first-order lags as elementary components,
but interconnected in different ways. 

In Lindeberg's time-causal, time-recursive construction, progressively
coarser representations are obtained by cascading in series the truncated
exponential kernels \cite{Lindeberg2016}: 
\begin{equation}
h_{\tau}(t)=\frac{1}{\tau}e^{-t/\tau}\mathbf{1}_{[0,\infty)}(t),\quad\tau>0.\label{eq:comparison-elementary}
\end{equation}
Writing $\ell_{0}=u$ and $\ell_{j}=h_{\tau_{j}}*\ell_{j-1}$, the
transfer function from $u$ to $\ell_{j}$ is 
\begin{equation}
Q_{j}(s)=\prod^{j}_{i=1}\frac{1}{1+\tau_{i}s}.\label{eq:comparison-cascade}
\end{equation}
Each additional stage diminishes sign variations, giving $S(\ell_{j})\leq S(\ell_{j-1})$.

The distinction is due to the different nature of past-to-future and
future-to-future operators. For a causal linear system with impulse
response $g$, the zero-state response to a future input $u_{+}$
is 
\begin{equation}
(T_{g}u_{+})(t)=\int^{t}_{0}g(t-s)u_{+}(s)ds,\qquad t\geq0,\label{eq:comparison-toeplitz}
\end{equation}
whereas the free future response generated by a past input $u_{-}$
is 
\begin{equation}
(H_{g}u_{-})(t)=\int^{0}_{-\infty}g(t-s)u_{-}(s)ds,\qquad t\geq0.\label{eq:comparison-hankel}
\end{equation}
Here $T_{g}$ and $H_{g}$ are the Toeplitz and Hankel operators,
respectively. Introducing past age $r=-s$ in \eqref{eq:comparison-hankel}
gives the kernel $g(t+r)$, in contrast to the difference kernel $g(t-s)$
in \eqref{eq:comparison-toeplitz}. Variation diminishing of these
two kernels thus leads to different interconnections. Such Toeplitz-Hankel
and series-parallel distinctions have already been established for
finite-dimensional discrete-time systems in~\cite{grussler2022variation}.

\section{Nonlinear Scale-Space Memory and Universal Approximation}

The linear theory developed in the previous section gives a rather
rigid picture. Once nonlinearities are allowed, this
representation is no longer expected to remain exact. Nevertheless,
first-order lags continue to appear as natural building blocks.

There are at least two simple ways in which nonlinearity can enter
a memory representation. The first is through a nonlinear readout
of linear memory coordinates.
\begin{example}
Consider a collection of first-order lags 
\[
\tau\dot{x}_{i}=-a_{i}x_{i}+a_{i}u,\quad0<a_{1}<a_{2}<\cdots<a_{n},
\]
followed by the nonlinear readout 
\[
y_{\tau}(t)=\phi\left(\sum^{n}_{i=1}w_{i}x_{i}(t)\right),
\]
where $w_{i}>0$ and $\phi:\mathbb{R}\to\mathbb{R}$ is strictly increasing,
satisfies $\phi(0)=0$, and 
\[
|\phi'(x)|\leq L,\qquad\forall x\in\mathbb{R},
\]
for some $L>0$.Let $L_{a_{i},\tau}$ denote the first-order lag whose
transfer function is 
\begin{equation}
\widehat{L}_{a_{i},\tau}(s)=\frac{1}{\tau s/a_{i}+1}.\label{eq:hat La tau}
\end{equation}
The corresponding whole-line input-output operator can then be written
as 
\begin{equation}
F_{\tau}=\phi\circ\left(\sum^{n}_{i=1}w_{i}L_{a_{i},\tau}\right).\label{eq:F1}
\end{equation}
The linear operator inside the parentheses is a finite-dimensional
linear scale-space memory. Since $\phi$ is strictly increasing and
$\phi(0)=0$, applying $\phi$ pointwise does not change the ordered
sign pattern of a signal. Consequently, the corresponding Hankel operator
remains ${\rm OVD}$. Scale covariance is also preserved because
the nonlinearity acts only on the amplitude and does not introduce
an additional temporal scale.
\end{example}

Nonlinearity need not be confined to the readout. It may also occur
inside the dynamics themselves, as the following example shows.
\begin{example}
Consider a nonlinear dynamics
\begin{align*}
\tau\dot{x} & =-x+u,\\
\tau\dot{z} & =-az+ax^{3},\quad a>0,\\
y & =z.
\end{align*}
Scale-covariance is automatic. The systems define an one-parameter
systems $F_{\tau}:L_{2}(\mathbb{R})\to L_{2}(\mathbb{R})\cap C_{b}(\mathbb{R})$
whose restrictions induce the Hankel operators $\{H_{\tau}\}_{\tau}>0$.
We check if $H_{\tau}\in{\rm OVD}$. Let $\widehat{L}_{a,\tau}=\frac{1}{\tau s/a+1}$
whose time-domain operator is denoted as $L_{a,\tau}$. Then
\begin{equation}
F_{\tau}=L_{a,\tau}\circ\phi\circ L_{1,\tau}\label{eq:F2}
\end{equation}
where $\phi(x)=x^{3}$. We have
\begin{align*}
S(F_{\tau}u) & =S(L_{a,\tau}\circ\phi\circ L_{1,\tau}u)\\
 & \le S(\phi\circ L_{1,\tau}u)\\
 & =S(L_{1,\tau}u)\\
 & \le S(\widetilde{u}).
\end{align*}
The same argument preserves the ordering of the signs whenever equality
holds. Hence the associated Hankel operators satisfy 
\[
S(H_{\tau}u)\le S(\widetilde{u})\implies H_{\tau}\in{\rm OVD}.
\]
\end{example}

The two examples suggest a useful principle. In the linear theory,
first-order lags are not merely convenient realizations but also the
elementary components forced by scale covariance and variation diminishing.
The examples above indicate that these same components remain useful
after nonlinearities are introduced, either through a nonlinear readout
or through their interconnection with nonlinear maps.

This leads to the central question of the nonlinear theory:
\begin{quote}
\emph{Can an arbitrary nonlinear scale-space memory be approximated,
to any desired accuracy, by finitely many first-order memory coordinates
followed by a static nonlinear readout? }
\end{quote}
This is a universal-approximation problem. In contrast with the linear
case, however, scale covariance and variation diminishing alone do
not provide enough regularity to obtain such a result. We must additionally
control how sensitively the memory depends on events that
occurred far in the past.

The appropriate notion is \emph{fading memory}. It was employed in
nonlinear system approximation by Boyd and Chua \cite{boyd1985fading}.
Informally, fading memory expresses the idea that the distant past
should matter less than the recent past: two histories that agree
approximately near the present should produce approximately the same
present response, even if they differ farther back in time.

In \cite{boyd1985fading}, the authors worked with signal spaces $C_{b}(\mathbb{R})$
with supremum norm (with a weighting function). We choose to work
with $L_{2}$ since we need to deal with truncated signals that are not continuous on $\mathbb{R}$. Let 
\[
w:[0,\infty)\to(0,1],\;w(0)=1,\;w(t)\xrightarrow{t\to\infty}0.
\]
Define $L_{2}$-norms on $\mathbb{R}_{-}$ as
\[
\|u\|_{w,\tau}=\left(\frac{1}{\tau}\int^{0}_{-\infty}w(-s/\tau)|u(s)|^{2}ds\right)^{1/2},\quad\tau>0.
\]
The normalization by $\tau$ is chosen so that the weighted topology
transforms naturally under time dilation. Indeed, 
\[
\|D^{-}_{c}u\|_{w,c\tau}=\|u\|_{w,\tau}.
\]
Thus the notion of fading memory itself respects the scale covariance
of the memory representation.

Define the aged admissible input set 
\begin{equation}
K_\tau^{\#}:=\{\Gamma_t u: u\in K_\tau: t\ge 0 \}\subset L_{2}(\mathbb{R}_{-}),
\end{equation}where $\Gamma_t u(s) = u(s+t)$ for $s\le t$ and $\Gamma_t u(s)=0$ otherwise.

For a given scale space memory $\{H_\tau, K_\tau \}_{\tau>0}$,
we impose the fading-memory on $K_\tau^\#$ rather than
$K_\tau$ because the universal approximation result concerns the
response uniformly over all future times. Indeed,
$H_\tau u(t)=\Phi_\tau(\Gamma_tu)$. Since aging may take admissible histories
outside $K_\tau$, continuity on $K_\tau$ alone does not suffice.
The set $K_\tau^\#$ therefore provides the natural domain for
the fading-memory condition required by the approximation theorem.

\begin{defn}
Let $\{H_{\tau},K_{\tau}\}_{\tau>0}$ be a nonlinear memory scale
space. We say that $\{H_{\tau},K_{\tau}\}_{\tau>0}$ has\emph{ $w$-fading
memory} if 
\[
\Phi_{\tau}:K_\tau^\# \to\mathbb{R},\quad\Phi_{\tau}(u)=H_{\tau}u(0)
\]
is continuous with respect to the weighted norm $\|\cdot\|_{w,\tau}$,
for all $\tau>0$.
\end{defn}

Equivalently, $\{H_{\tau},K_{\tau}\}_{\tau>0}$ has $w$-fading memory
if for every $\tau>0$, $u\in K^\#_{\tau}$ and $\epsilon>0$, there
exists $\delta=\delta(\tau,\epsilon,u)>0$, such that
\[
\forall v\in K^\#_{\tau},\;\|u-v\|_{w,\tau}<\delta\implies|H_{\tau}u(0)-H_{\tau}v(0)|<\epsilon.
\]

Consider the following past-input set
\begin{equation}
K_{1}=\{u\in H^{1}(\mathbb{R}_{-}):\|u\|_{L_{2}}\le M_{1},\;\|u'\|_{L_{2}}\le M_{2}\}\label{eq:setK}
\end{equation}
in which $H^{1}(\mathbb{R}_{-})$ is the Sobolev space on $\mathbb{R}_{-}$:
\[
H^{1}(\mathbb{R}_{-})=\{u\in L_{2}(\mathbb{R}_{-}):\|u'\|_{L_{2}}<\infty\}
\]
where $u'$ is the distributional derivative. 
To meet the definition of scale-space memory, $K_\tau$ is uniquely determined by $K_1$ (see Definition \ref{def:ss-memory}),
\begin{equation}
K_{\tau}=D^{-}_{\tau}K_{1}.\label{eq:K=00005Ctau}
\end{equation}
Thus $u\in K_{\tau}$ if and only if its normalized representation
$D^{-}_{1/\tau}u\in K_{1}$. 

Note that the set $K_{1}$ and $K_1^\#$ are not relatively compact in the ordinary
$L_{2}$ topology. They are, however, compact in the weighted $L_{2}$
topology. This parallels the construction of Boyd and Chua \cite{boyd1985fading},
who considered the signal class
{\small
\[
K=\{u\in C_{b}(\mathbb{R}):\|u\|_{\infty}<M_{1},\;\|T_{\tau}u-u\|_{\infty}\le M_{2}\tau,\,\forall\tau\ge0\}.
\]}Thus $K_{1}$ defined as \eqref{eq:setK} is the $L_{2}$ analogue of $K$ used in \cite{boyd1985fading}.

For the first order lag $\widehat{L}_{a,\tau}$ (see \eqref{eq:hat La tau}),
with $a>0,$ $\tau>0$, define 
\begin{equation}
x_{a,\tau}(u)=\frac{a}{\tau}\int^{0}_{-\infty}e^{as/\tau}u(s)ds.\label{eq:x a tau}
\end{equation}
We shall choose the weighting function $w$ so that $x_{a,\tau}$
is continuous in the weighted norm $\|\cdot\|_{w,\tau}$. In fact,
$w(r)=(1+r)^{-2}$ will do. 

For later use, we define the dead-zone function
\[
D_{\delta}(x)=\begin{cases}
x-\delta, & x>\delta,\\
0, & |x|\le\delta,\\
x+\delta, & x<-\delta
\end{cases}
\]
with $\delta\ge0$. 

\begin{thm} \label{thm:nonlinear}
Let $\{H_{\tau},K_{\tau}\}_{\tau>0}$ be a scale-space memory with
$w$-fading memory. Then for any $\epsilon>0$, there exist $n\in\mathbb{N}$,
$\delta\ge0$, $0<a_{1}<a_{2}<\cdots<a_{n}$, a polynomial $p\in\mathbb{R}[z_{1},\cdots,z_{n}]$,
and a scale-space memory $\{\overline{H}_{\tau},K_{\tau}\}_{\tau>0}$
with $w$-fading memory,
\[
\overline{H}_{\tau}u(t)=D_{\delta}\left(p(x^{1}_{\tau}(t),x^{2}_{\tau}(t),\cdots,x^{n}_{\tau}(t)\right),\quad t\ge0,
\]
in which 
\[
x^{i}_{\tau}(t)=\frac{a_{i}}{\tau}\int^{0}_{-\infty}e^{-\frac{t-s}{\tau/a_{i}}}u(s)ds,
\]
such that 
\begin{equation}
\sup_{\tau>0}\sup_{u\in K_{\tau}}\sup_{t\ge0}|\overline{H}_{\tau}(u)(t)-H_{\tau}(u)(t)|<\epsilon.\label{eq:sup tilde H H}
\end{equation}

Moreover, 
\begin{equation}
S(\overline{H}_{\tau}u)\le S(H_{\tau}u)\le S(\widetilde{u})\label{eq:S- tilde H}
\end{equation}
for every $u\in K_{\tau}$, $\tau>0$. If $S(\overline{H}_{\tau}u)=S(\widetilde{u})$,
then $\overline{H}_{\tau}u$ and $\widetilde{u}$ have the same sign
pattern.
\end{thm}

\begin{rem}
The above theorem now provides an affirmative answer to the universal-approximation question that we posed earlier in this section.
\end{rem}

\section{Conclusion and Future Research}

The paper presents a scale-space theory for memory. The main result is that two simple axioms, namely scale-covariance and order-preserving variation diminishing, effectively restrict memory to a narrow class of Hankel operators. In particular, linear scale-space memory is a parallel interconnection of first order lags, and nonlinear scale-space fading memory can be approximated by a linear one followed by a static nonlinear readout.

The next step is to develop an invariant geometry theory for scale-space memory and to apply them to the study of neuromorphic systems, in which memory plays a pivotal role.

\section{Proofs}

\subsection{Proof of Theorem \ref{thm:lin}}

Consider the reference scale $\tau=1$, and write $g:=g_{1}$ (see \eqref{eq:g_tau}).
After the change of variable $r=-s$, the Hankel operator becomes
\[
H_{1}(u)(t)=\int^{\infty}_{0}g(t+r)\widetilde{u}(r)dr,\quad\widetilde{u}(r):=u(-r).
\]
Thus $H_{1}$ is an integral operator with kernel $K(t,r)=g(t+r)$.
Since $H_{1} \in {\rm OVD}$, the classical variation-diminishing
characterization implies that $K$ is totally nonnegative of all orders.
By Karlin \cite[Chapter 2. Section 7-8, in particulr Theorem 7.3 and Lemma 7.4]{Karlin1968I},
there exists a nonnegative Borel measure $\nu$ such that
\[
g(t)=\int_{\mathbb{R}}e^{-at}d\nu(a).
\]
Since $g\in L_{2}(\mathbb{R}_{+})$ and $\nu$ is nonnegative, $\nu$
cannot charge $(-\infty,0)$, since otherwise negative $a$ would
produce exponentially growing terms. Hence
\begin{equation}
g(t)=\int^{\infty}_{0}e^{-at}d\nu(a).\label{eq:gt}
\end{equation}

It remains to show that the resulting measure is finite. Since the
whole-line extension $F_{1}$ is bounded on $L_{2}(\mathbb{R})$ and
$g\ge0$, $g$ is integrable, i.e., $\int^{\infty}_{0}g(r)dr<\infty$.
Using \eqref{eq:gt} and Tonelli's theorem \cite[Theorem 4.4]{brezis2011functional},
\[
\int^{\infty}_{0}g(r)dr=\int_{(0,\infty)}\frac{1}{a}d\nu(a)<\infty.
\]
We may therefore define $d\mu(a):=\frac{1}{a}d\nu(a)$. Then $\mu$
is a finite nonnegative Borel measure on $\mathbb{R}_{+}$, and
\begin{equation}
g_{1}(t)=\int_{(0,\infty)}ae^{-at}d\mu(a).\label{eq:g1}
\end{equation}

It remains to use scale covariance. By assumption,
\[
H_{c\tau}D^{-}_{c}=D^{+}_{c}H_{\tau},\;\forall c,\tau>0.
\]
Taking $\tau=1$ and $c=\tau$ gives
\[
H_{\tau}D^{-}_{\tau}=D^{+}_{\tau}H_{1}.
\]
Substituting the convolution representations and changing variables
yields
\[
\tau g_{\tau}(\tau s)=g_{1}(s),\quad\forall r\ge0,
\]
or equivalently $g_{\tau}(t)=\frac{1}{\tau}g_{1}(t/\tau)$. Combining
this identity with \eqref{eq:g1}, we obtain
\[
g_{\tau}(t)=\int_{(0,\infty)}\frac{a}{\tau}e^{-at/\tau}d\mu(a).
\]
Taking the Laplace transform gives
\[
G_{\tau}(s)=\int_{(0,\infty)}\frac{a}{\tau s+a}d\mu(a).
\]
The measure $\mu$ is independent of $\tau$, as claimed.

\subsection{Proof of Corollary \ref{cor:lin}}
The equivalence between 2) and 3) is obvious. 
Next, we prove 3) $\implies$ 1). For $\widehat{H}_{\tau}$, the impulse
response is 
\[
g_{\tau}(t)=\int_{(0,\infty)}\frac{a}{\tau}e^{-at/\tau}d\mu(a),\quad t\ge0
\]
and
\[
H_{\tau}u(t)=\int^{0}_{-\infty}g_{\tau}(t-s)u(s)ds,\quad t\ge0.
\]
For convenience, as before, denote $\widetilde{u}(t)=u(-t)$, $t\ge0$
and define
\[
\check{H}_{\tau}v(t):=\int^{\infty}_{0}g_{\tau}(t+s)\widetilde{u}(s)ds.
\]
Obviously, ${\rm rank}\check{H}_{\tau}={\rm rank}H_{\tau}$. Suppose
that $\widehat{H}_{\tau}$ has the form \eqref{eq:finite H}. Then
\[
g_{\tau}(t+s)=\sum^{n}_{i=1}c_{i}\phi_{i}(t)\phi_{i}(s)
\]
where $c_{i}=w_{i}a_{i}/\tau$ and $\phi_{i}(t)=e^{-a_{i}t/\tau}$.
Hence 
\[
\check{H}_{\tau}v(t)=\sum^{n}_{i=1}c_{i}\phi_{i}(t)\left\langle v,\phi_{i}\right\rangle _{L_{2}}.
\]
Note that $M_{ij}=\left\langle \phi_{i},\phi_{j}\right\rangle =\frac{\tau}{a_{i}+a_{j}}$
forms a positive definite matrix and hence invertible. Thus 
\[
\check{H}_{\tau}{\rm span}\{\phi_{1},\cdots,\phi_{n}\}={\rm span}\{\phi_{1},\cdots,\phi_{n}\},
\]
which implies ${\rm rank}H_{\tau}=n$ for all $\tau>0$. 

Finally, we prove 1) $\implies$ 3). We prove by contradiction. Assume
there exists $\tau>0$, such that $\widehat{H}_{\tau}$ has finite
rank and that $\mu$ has infinite support. For any $b>0$, define
$v_{b}(t)=e^{-bt}$. Then 
\begin{equation}
\check{H}_{\tau}v_{b}(t)=\int^{\infty}_{0}\frac{a}{\tau b+a}e^{-at/\tau}d\mu(a).\label{eq:tilde H}
\end{equation}
For any $m>0$ and a set of distinct constants $b_{1},\cdots,b_{m}>0$,
we show that $\{\check{H}_{\tau}v_{b_{i}}\}^{m}_{i=1}$ is linearly
independent. Indeed, if there exists some constants $\{c_{i}\}^{m}_{i=1}$
such that \[ \sum^{m}_{i=1}c_{i}\check{H}_{\tau}v_{b_{i}}=0, \]then
by \eqref{eq:tilde H}, we have
\[
\int^{\infty}_{0}e^{-at/\tau}d\eta(a)=0,\quad\forall t\ge0,
\]
in which 
\[
d\eta(a)=\left(\sum^{m}_{j=1}c_{j}\frac{a}{a+\tau b_{j}}\right)d\mu(a).
\]
Obviously $\eta$ is a finite signed Borel measure on $\mathbb{R}_{+}$.
By uniqueness of the Laplace transform for finite measures \cite[Proposition 1.2]{SchillingSongVondracek2012},
it follows that 
\[
\sum^{m}_{j=1}c_{j}\frac{a}{a+\tau b_{j}}=0\quad\mu\text{-a.e.}
\]
Let $\varphi(a):=\sum^{m}_{j=1}c_{j}\frac{a}{a+\tau b_{j}}$. Since
$\varphi$ is continuous, we have $R=0$ on ${\rm supp}\mu$. But
then $\varphi$ has infinitely many zeros which is impossible since
it is a finite order rational function. Thus, $\varphi(a)=0$ for
all $a>0$. Since $b_{j}$ are different, we deduce $c_{i}=0$ for
all $i=1,\cdots,m$. In other words, $\{\check{H}_{\tau}v_{b_{i}}\}^{m}_{i=1}$
is linearly independent. Since $m$ is arbitrary, this shows that
${\rm rank}\check{H}_{\tau}=\infty$, a contradiction.

\subsection{Proof of Theorem \ref{thm:nonlinear}}
The family $\{x_{a,1}:a>0\}$ separates points. Indeed, if $x_{a,1}(u)=x_{a,1}(v)$
for every $a>0$, then
\[
\int^{0}_{-\infty}e^{as}[u(s)-v(s)]ds=0,\quad\forall a>0.
\]
Uniqueness of Laplace transform gives $u=v$ a.e., \cite[Proposition 1.2]{SchillingSongVondracek2012}.
Therefore, on the compact set $K_\# := K_1^{\#}$, the algebra generated by
$\{1,x_{a,1}:a>0\}$ contains constants and separates points. Stone-Weierstrass
gives
\[
\forall\eta>0,\quad\exists n,a_{1},\cdots,a_{n}>0,\;p\in\mathbb{R}[z_{1},\cdots,z_{n}]
\]
such that
\[
\sup_{v\in K_{\#}}|\Phi_{1}(v)-p(x_{a_{1},1}(v),\cdots,x_{a_{n},1}(v))|<\eta.
\]
Substituting $v=\Gamma_{t}u$ into the above inequality and $H_{1}u(t)=\Phi_{1}(\Gamma_{t}u)$
and $x_{a,1}(\Gamma_{t}u)=e^{-at}x_{a,1}(u)$, it becomes 
\[
\sup_{u\in K_{1}}\sup_{t\ge0}|H_{1}u(t)-p(e^{-a_{1}t}x_{a_{1},1}(u),\cdots,e^{-a_{n}t}x_{a_{n},1}(u))|<\eta.
\]
By scale-covariance of $\{H_{\tau},K_{\tau}\}_{\tau>0}$, we obtain
\begin{equation}
\sup_{\tau>0}\sup_{u\in K_{\tau}}\sup_{t\ge0}|p(x^{1}_{\tau}(t),x^{2}_{\tau}(t),\cdots,x^{n}_{\tau}(t)-H_{\tau}(u)(t)|<\eta.\label{eq:esti-eta}
\end{equation}
Choose 
\[
0<\eta<\delta,\quad\eta+\delta<\epsilon
\]
and define
\[
\overline{H}_{\tau}u(t)=D_{\delta}(p(x^{1}_{\tau}(t),x^{2}_{\tau}(t),\cdots,x^{n}_{\tau}(t)).
\]
Since $|D_{\delta}(r)-r|\le\delta$, \eqref{eq:esti-eta} gives 
\[
\sup_{\tau>0}\sup_{u\in K_{\tau}}\sup_{t\ge0}|\overline{H}_{\tau}u(t)-H_{\tau}u(t)|<\delta+\eta<\epsilon,
\]
which is the desired inequality \eqref{eq:sup tilde H H}.

For variation diminishing, if $\overline{H}_{\tau}u(t)>0$, then
\[ 
p(x^{1}_{\tau}(t),x^{2}_{\tau}(t),\cdots,x^{n}_{\tau}(t)>\delta,
\]
as a consequence, by \eqref{eq:esti-eta}, 
\[
H_{\tau}(u)(t)>\delta-\eta>0.
\]
Similarly, $\overline{H}_{\tau}u(t)<0$ implies $H_{\tau}u(t)<0$.
Therefore, every alternating sign sequence of $\overline{H}_{\tau}u$
is also an alternating sign sequence of $H_{\tau}u$, with the same
signs at the same ordered times. Hence $\overline{H}_{\tau}\in{\rm OVD}(K_{\tau})$
and \eqref{eq:S- tilde H} holds. 

It is easy to see $\overline{H}_{c\tau}D^{-}_{c}=D^{+}_{c}\overline{H}_{\tau}$.
Therefore, $\{\overline{H}_{\tau},K_{\tau}\}_{\tau>0}$ is scale-covariant.

Finally, to see the $\overline{H}_{\tau}$ has $w$-fading memory,
note that 
\[
u\mapsto D_{\delta}(p(x_{a_{1},\tau}(u),\cdots,x_{a_{n},\tau}(u)))
\]
is continuous in the weighted $L_{2}$ norm since each $x_{a_{i},\tau}$
is, and that $p$ and $D_{\delta}$ are both continuous.

\bibliographystyle{IEEEtran}
\bibliography{main}

\end{document}